\documentclass[sigplan]{acmart}
\usepackage{algorithmic}
\usepackage{textcomp}
\usepackage{xcolor}
\usepackage{float}
\usepackage{afterpage}
\usepackage{changepage}
\usepackage{comment}
\usepackage{listings}
\usepackage{makecell}
\usepackage{svg}
\usepackage{tikz}
\usetikzlibrary{positioning}
\usepackage{csquotes} 
\usepackage{placeins}
\usepackage{soul}

\renewcommand\footnotetextcopyrightpermission[1]{}
\acmConference[PaPoC '25]{12th Workshop on Principles and Practice of Consistency for Distributed Data}{March 31, 2025}{Rotterdam, Netherlands - EuroSys/ASPLOS '25}

\include{commands/mathdefs}

\newcommand\centerxy[1]{
  \begin{center}
    \begin{tabular}{c}
      \begin{xy}
        #1
      \end{xy}
    \end{tabular}
  \end{center}
}

\auxfun{A}
\auxfun{B}
\auxfun{C}
\auxfun{D}
\auxfun{E}
\auxfun{BP}
\auxfun{RR}
\auxfun{m}
\auxfun{add}
\auxfun{receive}
\auxfun{send}
\auxfun{store}
\auxfun{operation}

\auxfun{inf}

\newcommand\HL[1]{\leavevmode\rlap{\hbox to \hsize{\color{#1}\leaders\hrule height .7\baselineskip depth .95ex\hfill}}}

\usepackage{nicefrac}
  
\usepackage{lipsum}
 
\usepackage{subcaption}

\usepackage[all]{xy} 

\usepackage{amsthm}
\theoremstyle{plain}

\usepackage{pifont}

\usepackage[noline,noend,linesnumbered]{algorithm2e}
\usepackage{multicol}

\usepackage{booktabs}

\title{CRDT-Based Game State Synchronization in Peer-to-Peer VR}

\author{Abel Dantas}
\email{adantas@ctvc.pt}
\affiliation{
  \institution{ProDEI, Universidade do Porto}
  \country{Portugal}
}

\author{Carlos Baquero}
\email{cbm@fe.up.pt}
\affiliation{
  \institution{Universidade do Porto \& INESC TEC}
  \country{Portugal}
}

\begin{document}

\begin{abstract}
Virtual presence demands ultra-low latency, a factor that centralized architectures, by their nature, cannot minimize. Local peer-to-peer architectures offer a compelling alternative, but also pose unique challenges in terms of network infrastructure.

This paper introduces a prototype leveraging Conflict-Free Replicated Data Types (CRDTs) to enable real-time collaboration in a shared virtual environment. Using this prototype, we investigate latency, synchronization, and the challenges of decentralized coordination in dynamic non-Byzantine contexts.

We aim to question prevailing assumptions about decentralized architectures and explore the practical potential of P2P in advancing virtual presence. This work challenges the constraints of mediated networks and highlights the potential of decentralized architectures to redefine collaboration and interaction in digital spaces.
\\

\end{abstract}

\maketitle

\section{Introduction}
Latency remains a fundamental obstacle in achieving immersion in VR. Existing cloud-based architectures introduce network delays that can disrupt user experience and exacerbate VR sickness. On the other hand, local peer-to-peer (P2P) collaboration promises ultra-low latency, but its adoption has been stifled by challenges like NAT traversal and limited support in public and semi-public networks. To the best of our knowledge, this paper represents the first exploration of Conflict-Free Replicated Data Types (CRDTs) within the context of VR.

We hypothesize that integrating P2P architectures with Conflict-Free Replicated Data Types will minimize latency in these environments.  This paper evaluates the impact of CRDT integration on user experience and its broader implications for the development of real-time, dynamic collaborative environments. This leads us to the following research questions:
\begin{itemize}
    \item Can CRDTs serve as a foundational technology for collaboration in P2P VR systems?
    \item What is the magnitude of network latency reduction achievable with P2P architectures?
\end{itemize}

We will thus focus on the technical evaluation of P2P architectures and CRDTs for VR collaboration; however, the broader implications of architectural paradigms applied to VR, particularly the contrast between centralized and decentralized systems, are hard to ignore. As virtual environments become central to work and social interactions, centralized networks, such as the Meta VR ecosystem, pose risks of restricted access and data exploitation~\cite{johnson2024metaVR}. Decentralized architectures are essential to safeguard equitable and open digital spaces. This paper explores the implementation and performance of these technologies, offering a foundation to address challenges and inform the design of future VR systems.

\section{Background and Related Work}

\subsection{Network Latency in VR}

Network latency is critical for collaborative VR systems, particularly with increasing synchronized data volumes. Photon-to-motion latency exceeding 100 ms significantly degrades user experience~\cite{casermanEffectsEndtoendLatency2019, anuaSystematicReviewPurpose2022, latoschikNotAloneHere2019}. Similarly, network delays over 230 ms impair task performance~\cite{becherNegativeEffectsNetwork2020} and disrupt shared presence in multiuser VR~\cite{slaterHowWeExperience2009}.

P2P networks can reduce latency and cost by leveraging user devices instead of centralized servers. In MMOGs (massively multiplayer online games), hybrid P2P-edge server approaches reduce latency by up to 45\%~\cite{plumbHybridNetworkClusters2018}. While P2P transmission strategies address bandwidth bottlenecks~\cite{huWebTorrentBasedFinegrained2017}, P2P systems face challenges with NAT traversal and firewalls~\cite{vanhardenbergPushPinProductionqualityPeertopeer2020}.

\subsection{Conflict-Free Replicated Data Types (CRDTs)}

CRDTs ensure eventual consistency through independent updates and conflict resolution without centralized coordination~\cite{almeidaApproachesConflictfreeReplicated2024}. They are categorized as state-based, with high communication overhead, and operation-based, which are bandwidth-efficient but depend on reliable delivery~\cite{shapiroConflictfreeReplicatedData2011}. Delta State CRDTs (\(\delta\)-CRDTs) combine these strengths by transmitting compact delta-states~\cite{almeidaDeltaStateReplicated2018}.

CRDTs are widely used in distributed systems, including Riak, Redis CRDBs, Azure Cosmos, and collaborative tools like Yjs and Automerge~\cite{kleppmannLocalfirstSoftwareYou2019}. Automerge, while capable of operating in P2P contexts, reflects skepticism about the practicality of pure P2P systems in consumer applications, favoring architecture-agnostic designs~\cite{vanhardenbergPushPinProductionqualityPeertopeer2020}. Similarly, Yjs excels in browser-based environments, but has not been optimized for dynamic, game-like VR settings, highlighting a gap for further exploration.

\section{Methodology}

We employed iterative prototype development to investigate latency and synchronization in VR collaboration. By \textbf{reframing latency as a data consistency challenge}, we explored CRDTs as a solution for decentralized collaboration. Each iteration addressed practical issues, such as peer-to-peer synchronization and conflict resolution, while evaluating their influence on real-time performance and the applicability of CRDTs in dynamic environments.

In this initial exploration, we focused primarily on the latency and effectiveness of CRDTs in a real-time virtual reality environment. Our approach proved to be robust against temporary network partitions. However, aspects such as synchrony conditions were not addressed within the scope of this study. The real-time nature of the application and the continuous broadcast of updates effectively mitigated many network-related challenges, resulting in a seamless user experience. Investigating the system's behavior under more severe conditions, such as prolonged network partitions or significant message loss, could offer deeper insights into how consistency and reliability manifest in this type of architecture.

\section{BrickSync: VR P2P CRDTs in Action}

BrickSync is an initial exploration of a framework for CRDT-based game-state synchronization, implemented within the Unity game engine, which allows us to collaborate in VR without a server.

\begin{figure}[h!]
    \centering
    \includegraphics[width=0.8\linewidth]{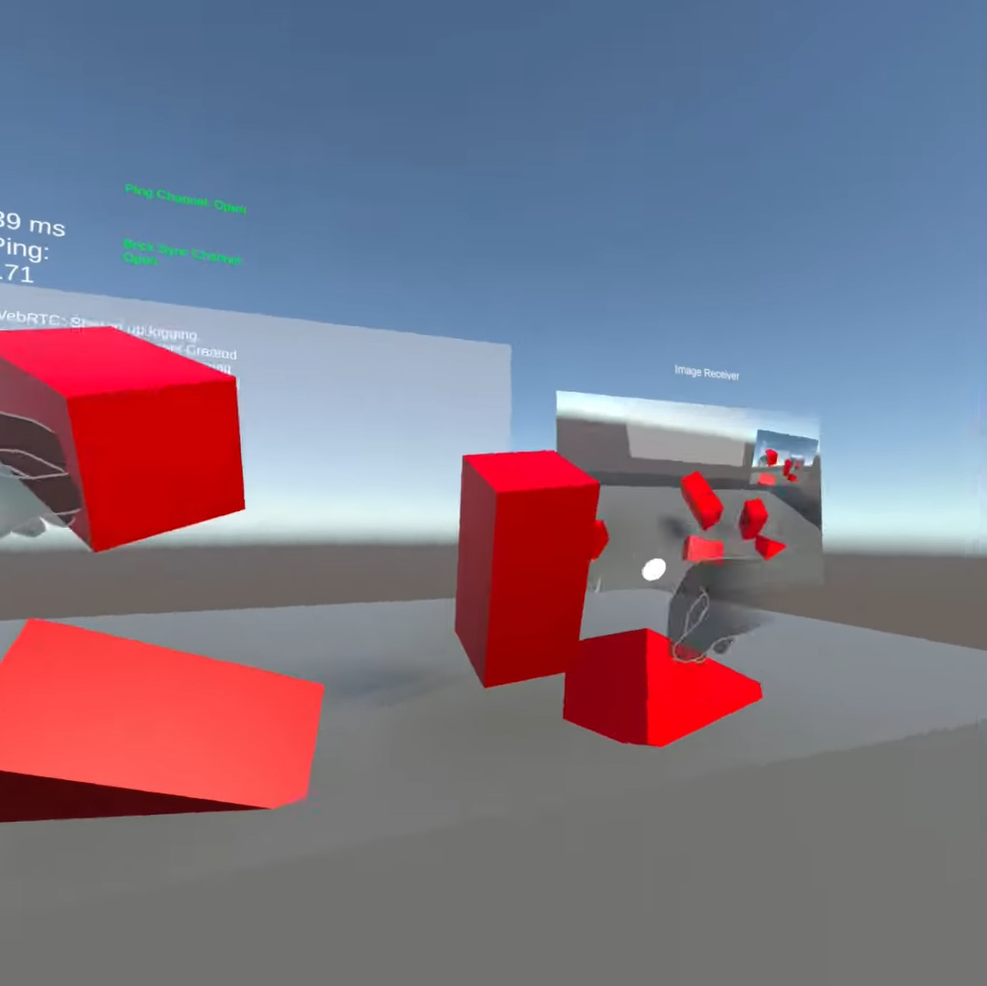}
    \caption{A snapshot of BrickSync in action, showcasing two users collaboratively manipulating virtual objects in a shared VR environment. Real-time synchronization of object states is achieved using CRDTs over a P2P WebRTC connection.}
    \label{fig:methods}
\end{figure}

Users interact with the environment using standard Meta Quest 3 controls, with support for both hand tracking and controller input. This allows users to grab and manipulate bricks using one or both hands.

Users collaboratively manipulate bricks in a shared environment, and their interactions are synchronized through CRDTs. The virtual scene consists of a table, a button for spawning bricks, and the ability to collaboratively build structures by moving and aligning bricks.

To assist in connection monitoring, a fixed position video feed window hovers near the main interaction area. This real-time feed displays what other users are seeing. In VR, where the headset obscures the view-port, this feature helps participants understand the actions of others and effectively assess the connection status.

Additional debugging and control tools include: real-time ping monitoring, an average ping record over the last minute, a connection status display for all WebRTC data channels, and a window for viewing local logs.

\subsection{Communication and Networking}

In BrickSync, the communication protocol has 2 elements:
\begin{enumerate}
    \item \textbf{Peer Discovery}: A WebSocket signaling server is used to establish connections between peers. Once connected, the signaling server is no longer involved in data exchange between peers.
    \item  \textbf{P2P Channels}: WebRTC data channels are used for real-time communication between peers. These channels transmit CRDT updates, enabling synchronized state management across all connected devices.
\end{enumerate}

\subsubsection{Peer Discovery}

The peer discovery process begins with one peer creating an SDP (Session Description Protocol) offer and sending it to another peer via a signaling server. The receiving peer responds with its own SDP, and both exchange ICE (Interactive Connectivity Establishment) candidates to establish a direct WebRTC connection. Once the connection is established, the signaling server is no longer involved, serving only to facilitate the initial handshake and maintain awareness of peer activity.

We recognize that using a signaling server deviates from a pure peer-to-peer (P2P) setup. This choice was driven by time and scope constraints, as implementing fully decentralized P2P signaling was beyond the objectives we set out to achieve. The challenge of bootstrapping nodes in a peer-to-peer network has long been recognized. For example, Kademlia assumes that at least one known node is needed for initialization~\cite{maymounkovKademliaPeertoPeerInformation2002}.

In general, research has shown that the bootstrap problem often necessitates trade-offs between decentralization, scalability, and complexity. For example, leveraging auxiliary systems like public overlays provides practical solutions to bootstrap peers under constraints such as NAT traversal, but often requires additional infrastructure and operational overhead ~\cite{wolinskyAddressingP2PBootstrap2010}. Other approaches, such as employing distributed mechanisms like DNS or DHT entries, rely on peers maintaining accurate and up-to-date information about the network~\cite{knollBootstrappingPeertoPeerSystems2008}. While alternative methods like Bluetooth or blockchain-based mempool signaling could theoretically eliminate server reliance, they would fundamentally shift the focus and complexity of this work.

Instead, we opted for a lightweight signaling server, deployed locally and remotely on Azure at zero cost. The server’s sole responsibility is to facilitate peer discovery, without influencing or participating in the subsequent data exchange.

\subsubsection{P2P Communication}

Our WebRTC data channels rely on SCTP (Stream Control Transmission Protocol) and support both ordered and unordered delivery. For CRDT synchronization, we configured the channel for ordered delivery, achieving source FIFO (First In, First Out) order.



Previous research has explored WebRTC for peer-to-peer collaborative editing in browsers~\cite{nedelec2016crate}, highlighting its potential despite early limitations on mobile devices. However, since then WebRTC mobile support has expanded, making it a viable option for real-time communication on smartphones. This is particularly relevant for VR applications on devices like the Meta Quest 3, which runs on Android and supports WebRTC. Thus, our choice of WebRTC aligns with the growing capabilities of mobile VR platforms.

The local RTC peer connections were also initialized with the option to use a STUN server, such as Google's public STUN servers. This facilitates the discovery of public IP addresses and port mappings, resolving NAT traversal issues in site-to-site communication. This is particularly useful in semi-public networks, such as academic environments like Eduroam~\cite{eduroam}, where peers are often behind NATs that block direct connections~\cite{rfc5389}. Although not strictly needed in controlled environments, the use of a STUN server is a valuable tool for testing in heterogeneous network configurations.

Each peer is also configured to handle incoming messages from other peers. This includes rendering the video channel to the video feed window and relaying data channel messages to the CRDT logic.

To optimize communication, we implemented separate data channels: one for Round-Trip Time (RTT) measurements and another for CRDT updates. The RTT channel tracks latency, while the CRDT channel is used for the synchronization of the in-game bricks.

\subsection{CRDT Integration}

Our CRDT implementation evolved through multiple phases. We first explored a naive operation-based approach and later a state-based approach. In our implementation, a Unity \texttt{MonoBehaviour} operates as adapter~\cite{gof_design_patterns} between the representation in the game engine and the CRDT. A local map per replica is also used to keep track of all the bricks -- and the corresponding CRDTs -- as they are spawned.

\subsubsection{Operation based approach}

We initially chose an operation-based approach with the goal of reducing communication overhead while maintaining consistent, frame-by-frame updates. In this approach, each update is represented as an \textit{operation} broadcast to all replicas, which then apply the operations. To accommodate users' real-time, continuous interactions with bricks, translation offsets are calculated and transformed into discrete operations.

In order to ensure that each operation is transmitted maintaining causal order, we used Vector Clocks to track the order of operations, allowing us to determine whether an incoming operation has already been applied, is not yet ready to be applied (due to unmet dependencies), or is the next in the logical sequence. In the latter case, the operation can be applied to the local replica.

Although operation-based CRDTs can perform well for real-time interactions over reliable channels, in practice, reliable channels are hard to come by and can be a source of extra latency. While the usage of vector clocks together with the FIFO ordering afforded by WebRTC proved sufficient for a two-user interaction, to achieve reliability beyond point-to-point or small groups, a group membership mechanism is necessary to properly handle view changes. Tools like Spread~\cite{spreadtoolkit} can be used to provide reliable causal ordering and possibly overcome the need for vector clocks as ordering would be relegated to the middleware. 

Causal delivery, either via specialized middleware or by application level ordering with vector clocks, might deliver concurrent operations in different order at different replicas. This requires special handling at the application layer unless these operation are already commutative.  
 

\begin{figure}[h!]
    \centering
    \includegraphics[width=\linewidth]{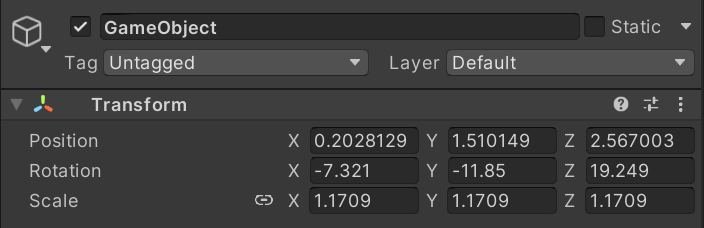}
    \caption{A Unity GameObject Transform component. Rotation is non-commutative: the order of rotations affects the final orientation. Rotation is a quaternion (4D space to prevent gimbal lock) represented in Euler angles on the UI. }
    \label{fig:quaternions}
\end{figure}

For example, in BrickSync, non-commutative operations -- such as rotating an object as highlighted in Figure~\ref{fig:quaternions}, can lead to divergent outcomes if applied in different orders. 
Within operation-based CRDTs, this issue can be tackled using one of two strategies:
\begin{itemize}
    \item Canonical Ordering: For example a Last-Writer-Wins (LWW) approach, which would avoid surprise merges by making one rotation “win” for concurrent actions. UX is not truly collaborative.
    \item Specialized Handling: For example averaging the rotations or accumulating deltas. This is more collaborative, but, might feel non intuitive if the final orientation is unexpected.
\end{itemize}

These approaches to handling non-commutative operations are inherently limited, and expose another drawback of operation-based CRDTs: without a comprehensive chronicle of the object, we are constrained on the strategies for reconciling conflicts. Although operation-based CRDTs still operate on a locally maintained state, and this local state could theoretically support techniques like representing divergent views using metadata, doing so would complicate the model and blur the boundaries of what defines an operation-based CRDT~\cite{baquero2024}. State-based CRDTs, on the other hand, propagate the entire state and, in this case, we found that they provide a better framework for ensuring an adequate handling of commutativity.



\newcommand{\stateA}{\alpha}
\newcommand{\stateB}{\beta}
\newcommand{\stateAB}{\{\alpha,\beta\}}
\newcommand{\stateBPrime}{\beta'}
\newcommand{\stateFinal}{\{\alpha,\beta'\}}

\begin{figure*}[!htb]
  \centerxy{
\xymatrix@C=2em @R=2em{
\A &
  \varnothing \ar@{->}[rr]^(0.45){\text{Move Right}} & & \stateA \ar@{}|{\raisebox{4em}{\text{\scriptsize \( \stateA = A_{\text{view}}[x_{1\times0}] = (1,0,0) \)}}}
  \ar@{.}[rr] & & \stateAB \ar@{}|{\hspace{12em}\raisebox{4em}{\text{\scriptsize \( \stateAB = \{A_{\text{view}}[x_{1\times0}], B_{\text{view}}[x_{0\times1}]\} = (0,0,0) \)}}}
  \ar@{.}[r] & \bullet^2
  \ar@{->}[rd]^(.6){\stateAB}
  \ar@{.}[rrr] & & & \stateFinal \ar@{}|{\hspace{10em}\raisebox{-7em}{\text{\scriptsize \( \stateFinal = \{A_{\text{view}}[x_{1\times0}], B_{\text{view}}[x_{0\times2}]\} = (-1,0,0) \)}}}
\\
\B &
  \varnothing \ar@{->}[rr]^(0.45){\text{Move Left}} & & \stateB \ar@{}|{\raisebox{-4em}{\text{\scriptsize \( \stateB = B_{\text{view}}[x_{0\times1}] = (-1,0,0) \)}}} 
  \ar@{.}[r] & \bullet^1
  \ar@{->}[ru]_(.65){\stateB}
  \ar@{.}[r] & \ar@{->}[r]_(0.45){\text{Move Left Again}} & \stateBPrime \ar@{}|{\raisebox{-4em}{\text{\scriptsize \( \stateBPrime = B_{\text{view}}[x_{0\times2}] = (-2,0,0) \)}}} 
  \ar@{.}[r] & \bullet^3
  \ar@{->}[ru]_(.65){\stateFinal}
}
}
\caption{MV-Transformer synchronization in Local-Space Mode with replicas A and B. A applies 'Move Right' to reach (1,0,0); B applies 'Move Left' twice to (-2,0,0). Solid arrows show local updates, dotted lines indicate state exchanges. The process merges to a final state of (-1,0,0).}
\label{fig:mv-transformer-diagram}
\end{figure*}
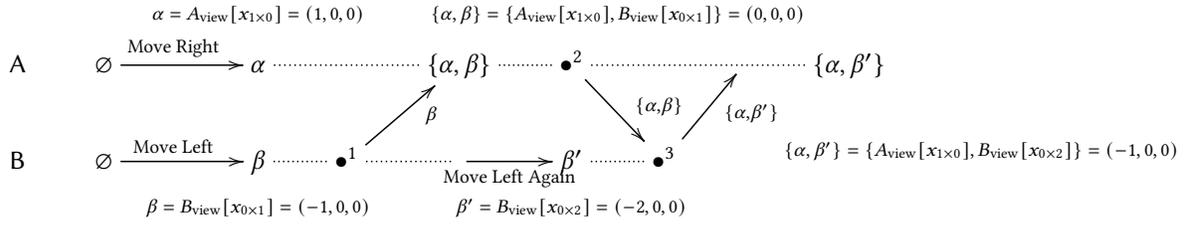

As we will see next, state-based CRDTs allow replicas to reconcile divergent states by encoding discrepancies in auxiliary properties -- additional metadata or dimensions separate from the conflicting attributes. In the context of real-time dynamic collaboration, this approach—at the expense of increased bandwidth and storage requirements—supports a broader range of conflict resolution strategies, including prioritization, blending, or heuristic-driven reconciliation.

\subsubsection{State based approach}

Faced with the aforementioned limitations of operation-based CRDTs -- (1) reliance on causal order and commutative operations and (2) the lack of a full chronicle of the object (the object's complete state history, including all changes and associated metadata) -- we opted to implement a state-based approach. Based off Unity's native data structures, we developed a MV-Transformer (Figure~\ref{fig:mv-transformer-diagram}), a higher-level state-based CRDT designed to encapsulate and synchronize a GameObject's Transform state. In addition to mediating state resolution for position, rotation, and scale, the MV-Transformer incorporates a register for tracking which replica(s) are currently manipulating the object -- in other words, which users are holding on to the object using the Meta Quest 3 controls. 
The MV-Transformer was engineered to toggle between two synchronization strategies for the underlying vectors: local-space updates and world-space updates. The local-space strategy, inspired by PN-Counters -- a type of CRDT that track increments and decrements separately, allowing for consistent concurrent updates to numerical values -- applies synchronization based on offset-based updates relative to the object's last known position. In contrast, the world-space strategy, modeled after a Last-Writer-Wins (LWW) approach, synchronizes using the latest absolute position of the object in world space. Instead of relying on a traditional feature flag to toggle between these strategies, the MV-Transformer also contains an internal boolean register that can dynamically dictate the synchronization mode (local-space or world-space).


To clarify the difference between the synchronization strategies, consider \textbf{world-space mode}. When multiple users interact with the same object in world-space mode, the object appears to oscillate between their control. Once one of the users releases their grip, the object returns to the hand of the other user -— the one who hold on to the object the longest. This simple method of conflict resolution is effective in scenarios where users are collaborating and actively communicating, as they can quickly recognize the conflict and coordinate who should release the object.

In \textbf{local mode}, on the other hand, synchronization relies on offset-based updates relative to the object's last known position. This approach prevents oscillations but may require more sophisticated handling of positional drift between replicas.

Through this prototype, we found that the effectiveness of CRDT-based architectures hinges on how divergences — such as conflicting updates or superimposed states — are presented to users. Providing clear, consistent, and navigable representations of these inconsistencies is crucial for collaboration.

\section{Findings and Discussion}

This section discusses key findings and challenges from the prototype development, focusing on technical and UX observations from the development process (A), conflict resolution in collaborative interactions in VR (B) and performance and latency analysis (C).

\subsection{Technical and UX Observations}

The development process revealed several technical and UX challenges.

\subsubsection{CRDT Developer Experience}
The elegance of CRDTs made implementing the shared state intuitive. Synchronization occurred naturally, creating the illusion of a shared, independent space. Testers often expressed surprise at the absence of a centralized server, and that you can conduct the experience with just two VR headsets -- even without an internet connection. We interpret this as a positive indication that CRDTs are a natural fit for architecting shared game-state in collaborative systems. All other technical challenges relate to establishing and maintaining the P2P connections.
\subsubsection{Heterogeneous Network Behavior}
Different network environments yielded varying behaviors. Testing in the semi-public networks caused erratic system behavior, creating confusion. This was mitigated with:
\begin{itemize}
    \item Resorting to a mobile hotspot for a stable testing environment.
    \item Hosting a remote WebSocket server on Azure for consistent signaling.
    \item Using a STUN server to address NAT traversal issues.
\end{itemize}
This experience highlights pain points in P2P systems when deployed to some networks and underscores the need for controlled network configurations to ensure consistent performance.

\subsubsection{Closing Data Channels}
WebRTC data channels closed arbitrarily on Meta Quest 3 devices, whereas this behavior was not observed during PC testing. The issue was mitigated by implementing a renegotiation strategy to reopen channels when they closed unexpectedly. While effective, this approach introduced additional latency and required specifying channel IDs during initialization. We speculate that the behavior stems from Android security policies or specific constraints within Meta Quest devices.

Such erratic channel closures and the resulting variability in delivery order have implications for idempotency and retransmission decisions. To address these challenges, state-based CRDTs offer robustness by ensuring consistency even under unpredictable network conditions or device-specific issues like those observed with Meta Quest 3.

Regarding fault tolerance, our system leverages the inherent properties of CRDTs to handle network partitions. Replicas can operate independently and merge their states upon reconnection, ensuring eventual consistency. For scenarios with high packet loss, the state-based CRDT approach provides robustness by allowing full state retransmission, while the ordered delivery of WebRTC helps maintain consistency for operation-based CRDTs. However, we assume a non-Byzantine environment, meaning malicious peers are not accounted for in the current design. Addressing such adversarial scenarios would require additional security mechanisms, which are beyond the scope of this initial exploration. While CRDTs provide a solid foundation for fault tolerance, extreme edge cases necessitate additional research.


\subsection{Global Rules}

In this section, we introduce the term \textit{Global Rules} to describe any state transformation that is not directly authored by a specific user. These transformations can be continuous -- such as gravity, which requires real-time updates -- or discrete -- such as a wind direction change or the spawning of an enemy in a game. They may also involve complex causal chains. For example, a user releases an object under the influence of gravity, which collides with a stack of dominoes initiating a chain reaction.

We refer to these as \textit{Global Rules} rather than environmental or physics rules, as they encompass any game logic capable of altering the game state without a specific author. While this touches on broader challenges in non-server-authoritative game engine design rather than solely state synchronization with CRDTs, we believe this perspective can inform the application of CRDTs in this context.

\subsection{Conflict Visualization}

When two users manipulated the same object simultaneously under world-space mode, the object oscillated between positions as the CRDT logic attempted to reconcile superimposed states. While this behavior did not significantly affect the user experience (on the contrary), it did raise questions about conflict resolution in collaborative contexts in general.

In the case of BrickSync we identified several strategies to address the issue with oscillation:
\begin{itemize}
    \item Prioritizing the most recent update (last-writer-wins).
    \item Introducing heuristics to maximize collaborative convergence, such as favoring positions that align with shared goals like building a straight wall.
    \item Averaging positions. In this case if two users manipulate the same object in different directions it should stand in the middle.
    \item Applying constraints to prevent simultaneous manipulations -- this one is a last resort.
\end{itemize}

Building on this, we propose embedding Dynamic Strategy Switching directly within CRDTs. This involves designing the CRDT to adaptively transition between conflict resolution strategies based on contextual factors such as interaction patterns, user intent, or system state. For example:

The CRDT could default to heuristic-based reconciliation, or last-writer-wins during cooperative tasks but switch to constraint-based mechanisms in competitive or high-conflict scenarios.
Thresholds or triggers, such as the frequency of conflicting updates or task-specific goals, would determine when and how the strategy changes.

Other heuristic-based conflict resolution strategies could employ a prediction layer, for example applying dead reckoning principles -- a method commonly used in fast-paced multiplayer games to estimate future positions of moving objects. By integrating this layer into the CRDT merge process, states aligning with the predicted path of an object could be prioritized to promote a smoother experience.

Heuristic-based approaches are directly applicable to existing frameworks such as Automerge or Yjs. For example, in a collaborative editing scenario, if two users are making conflicting contributions to the same text, a semantic heuristic could be applied to arbitrate, prioritizing the changes that align more closely with the document’s intent or structure. Alternatively, edits could be merged contextually based on their relevance to the overall goal, such as prioritizing keywords or stylistic consistency. This will become increasingly relevant as collaborative editing systems evolve to include real-time AI agents, where clear feedback and effective merging strategies for individual contributions are key.

\begin{figure*}[ht]
    \centering
    \includegraphics[width=\textwidth]{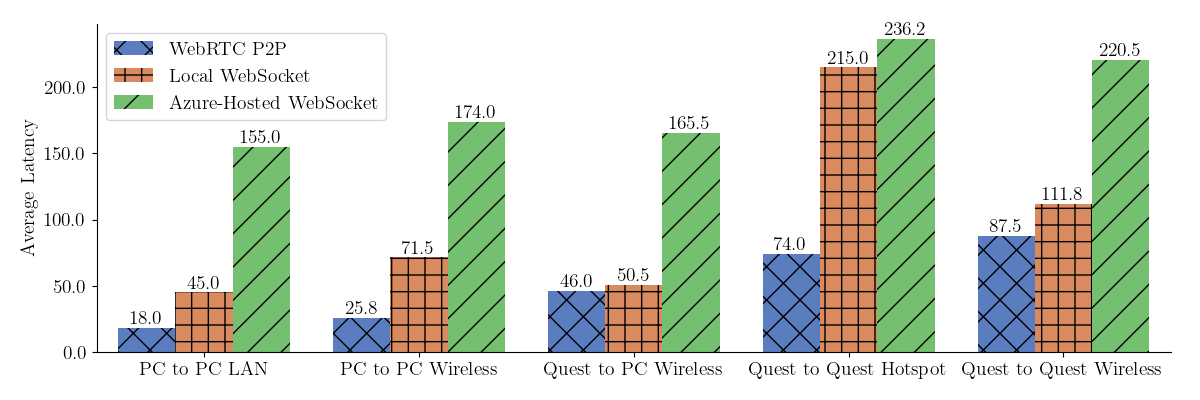}
    \caption{Impact of Connection Type and Architecture on Latency. The x axis displays different network configurations groups that show the different architectures, as expected, P2P is much faster.}
    \label{fig:average_latency_by_connection_type_and_architecture}
\end{figure*}

\subsection{Performance and Latency Analysis}

\begin{figure}[H]
    \centering
    \includegraphics[width=.45\textwidth]{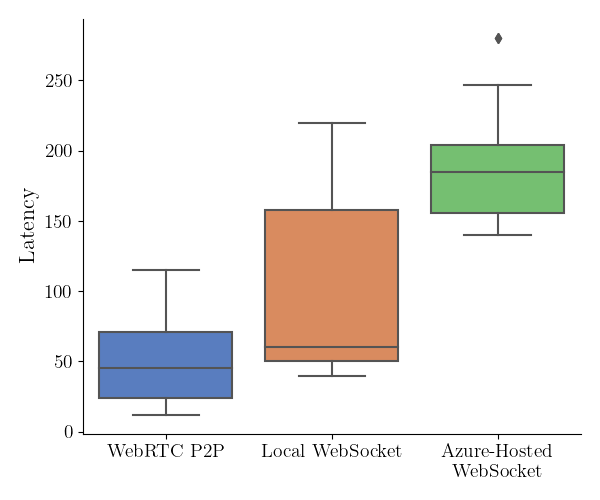}
    \caption{Distribution of latency results based on underlying architecture.}
    \label{fig:latency_distribution_by_architecture}  
\end{figure}

As part of our experiment with BrickSync, we collected latency data. Network latency was measured across various configurations, as shown in Table~\ref{fig:latency-measurements} and illustrated in Figure~\ref{fig:average_latency_by_connection_type_and_architecture} and ~\ref{fig:latency_distribution_by_architecture}.


While delays were noticeable in the WebRTC video stream, brick interactions remained fluid due to:
\begin{enumerate}
    \item Real-time CRDT updates providing immediate feedback for local actions.
    \item Users' inability to distinguish between network delay and the natural delay of other users' actions.
\end{enumerate}
By prioritizing local updates, CRDTs \textbf{minimize perceptible latency and mitigate VR sickness} caused by visual disconnects between user actions and system feedback.


While our experiments demonstrate the feasibility of CRDT-based synchronization in a two-user scenario, we acknowledge that this does not reflect the scalability required for real-world applications. The current setup was constrained by resources and aimed at initial exploration. Moreover, the choice of CRDT type has implications for scalability. State-based CRDTs, while simpler to implement, can lead to increased bandwidth usage due to the transmission of full states. Operation-based CRDTs, on the other hand, require reliable broadcast mechanisms, which can be complex to manage. Additionally, the use of vector clocks for maintaining causal order can result in significant memory and message size overhead as the number of replicas grow. These challenges were not pronounced in our two-user setup but need to be addressed, potentially through the adoption of delta-based CRDTs or other optimized approaches.

\begin{figure}[H]
    \centering
    \resizebox{\linewidth}{!}{%
    \begin{tabular}{|l|c|c|c|}
    \hline
    \textbf{Connection Type}      & \makecell{\textbf{WebRTC} \\ \textbf{P2P}} & \makecell{\textbf{Local} \\ \textbf{WebSocket}} & \makecell{\textbf{Azure-Hosted} \\ \textbf{WebSocket}} \\ \hline
    PC to PC - Ethernet             & 18 ms              & 45 ms                   & 155 ms                          \\ \hline
    PC to PC - WiFi                 & 25.75 ms           & 71.5 ms                 & 174 ms                          \\ \hline
    Quest to PC - WiFi              & 46 ms              & 50.5 ms                 & 165.5 ms                        \\ \hline
    Quest to Quest - WiFi           & 87.5 ms            & 111.75 ms               & 220.5 ms                        \\ \hline
    Quest to Quest - Hotspot        & 74 ms              & 215 ms                  & 236.25 ms                       \\ \hline
    \end{tabular}
    }
    \caption{Average latency measurements for different connection types and configurations across all test scenarios.}
    \label{fig:latency-measurements}
\end{figure}

It's also worthwhile to note that in scenarios with high number of replicas vector clocks become inefficient due to their size, that scales linearly with the number of nodes, resulting in significant metadata and communication overhead. Delta-state CRDTs can reduce bandwidth by transmitting only incremental updates, but still require end-to-end mechanisms to guarantee causal consistency.

Furthermore, we observe that empirical user studies should be conducted to validate the collaborative experience, perceived latency, and conflict resolution mechanisms. Understanding user feedback, particularly regarding phenomena like object oscillation in world-space mode, is crucial for refining the system. Our priority in this initial exploration was to establish technical feasibility and user feedback was sought but collected in a limited way.

In addition to these measurements, we conducted further tests to evaluate the impact of payload size. Unfortunately, WebRTC imposes a maximum payload size of 16 KB. And although comparative results show promising trends, to fully understand the implications, it would be necessary to implement message chunking to accommodate payloads exceeding the 16 KB limit, and conduct more tests to reduce the impact of outliers and network conditions.


\section{Conclusions}

This study demonstrates the feasibility of using CRDTs for decentralized VR collaboration within a P2P architecture. Our prototype demonstrates the ability to achieve low-latency, synchronized interactions without reliance on centralized servers. The controlled co-located VR environment served as a practical testbed to consider latency in a CRDT-based architecture.

Our findings demonstrate a substantial reduction in network latency when using P2P architectures compared to remote server setups. Specifically, the average latency in P2P configurations is approximately 50 ms, a 75\% reduction from the 200 ms observed in remote server connections. Moreover, P2P achieved the lowest recorded latency at 18 ms, compared to 45 ms on WebSocket-based setups, showcasing its ability to outperform other architectures by a factor of 2 in optimal conditions. While the use of more performant and well-connected remote servers may mitigate these differences, such configurations often incur higher costs and dependency on centralized infrastructure. P2P emerges as the most cost-effective solution for reducing latency, aligning the low-latency requirements of immersive VR environments with decentralization.

The interaction between CRDTs and Global Rules, such as gravity, revealed opportunities in handling non-user-authored state, revealing the importance of integrating adaptive conflict resolution strategies, such as dynamic strategy switching, directly into CRDTs. The findings suggest that decentralized systems must address these dynamics to support emergent behaviors in collaborative environments.

We further identified key challenges and opportunities in extending CRDTs and P2P systems to game-like, dynamic environments, particularly in conflict arbitration and scalability. We invite researchers and practitioners to build on these findings, exploring broader applications of decentralized architectures in VR and other collaborative settings. The potential of these systems to redefine collaboration and interaction in digital spaces is immense, and their future development holds promise for transformative impacts across industries and scientific domains.

\section{Future Work}

Our exploratory work revealed several promising directions to advance decentralized realtime collaborative environments, including local-first principles applied to VR and proximity-aware network topologies for in-loco coordination. However, the most critical avenue lies in embedding dynamic strategies, such as switching between heuristics and constraints based on interaction context, as discussed in Section V. Specifically, we hope to continue exploring how local-first CRDT designs can handle non-user-authored changes, such as physics-based interactions or other global rules. 

Scalability remains a critical area for further research. Experiments with a larger number of users and objects will be essential to understand the system's performance in more complex VR environments.

As decentralized architectures challenge centralized models, these efforts could pave the way for new interaction paradigms, positioning CRDTs as foundational building blocks for the open digital commons.

\section{Acknowledgments}

This work was partially financed by National Funds through the Portuguese funding agency, FCT (Fundação para a Ciência e a Tecnologia) under project LA/P/0063/2020 \--- DOI 10.54499/LA/P/0063/2020\footnote{https://doi.org/10.54499/LA/P/0063/2020}. We also thank the PRODEI Doctoral Program (Programa Doutoral em Engenharia Informática) at the Faculty of Engineering, University of Porto, and CTVC (Cooperativa Tecnológica de Viana do Castelo, a tech cooperative in northern Portugal) for their support and collaboration.


\newpage
\bibliographystyle{ACM-Reference-Format} 
\bibliography{References}   


\appendix
\section{Additional Data and Figures}

Below are additional figures and detailed latency measurements across all test scenarios. As well as some implementation and testing details.
\begin{figure}[h]
    \centering
    \resizebox{\linewidth}{!}{%
    \begin{tabular}{|l|l|l|c|c|c|}
    \hline
    \textbf{Date} & \textbf{Scenario} & \textbf{Connection Type} & \makecell{\textbf{WebRTC} \\ \textbf{P2P}} & \makecell{\textbf{Local} \\ \textbf{WebSocket}} & \makecell{\textbf{Azure-Hosted} \\ \textbf{WebSocket}} \\ \hline
    10-01-2025 & Minimal Payload & PC to PC LAN & 20 ms & 45 ms & 180 ms \\ \hline
    10-01-2025 & Minimal Payload & PC to PC Wireless & 25 ms & 60 ms & 140 ms \\ \hline
    10-01-2025 & Minimal Payload & Quest to PC Wireless & 50 ms & 47 ms & 150 ms \\ \hline
    10-01-2025 & Minimal Payload & Quest to Quest Wireless & 70 ms & 180 ms & 220 ms \\ \hline
    10-01-2025 & Minimal Payload & Quest to Quest Hotspot & 75 ms & 200 ms & 280 ms \\ \hline
    14-12-2024 & Minimal Payload & PC to PC LAN & 12 ms & 45 ms & 140 ms \\ \hline
    14-12-2024 & Minimal Payload & PC to PC Wireless & 22 ms & 101 ms & 158 ms \\ \hline
    14-12-2024 & Minimal Payload & Quest to PC Wireless & 34 ms & 55 ms & 177 ms \\ \hline
    14-12-2024 & Minimal Payload & Quest to Quest Wireless & 55 ms & 150 ms & 247 ms \\ \hline
    14-12-2024 & Minimal Payload & Quest to Quest Hotspot & 41 ms & 220 ms & 280 ms \\ \hline
    08-01-2025 & 16kb Payload & PC to PC LAN & 20 ms & 40 ms & 150 ms \\ \hline
    08-01-2025 & 16kb Payload & PC to PC Wireless & 30 ms & 60 ms & 200 ms \\ \hline
    08-01-2025 & 16kb Payload & Quest to PC Wireless & 50 ms & 50 ms & 170 ms \\ \hline
    08-01-2025 & 16kb Payload & Quest to Quest Wireless & 110 ms & 57 ms & 215 ms \\ \hline
    08-01-2025 & 16kb Payload & Quest to Quest Hotspot & 90 ms & 220 ms & 190 ms \\ \hline
    14-12-2024 & 16kb Payload & PC to PC LAN & 20 ms & 50 ms & 150 ms \\ \hline
    14-12-2024 & 16kb Payload & PC to PC Wireless & 26 ms & 65 ms & 198 ms \\ \hline
    14-12-2024 & 16kb Payload & Quest to PC Wireless & 50 ms & 50 ms & 165 ms \\ \hline
    14-12-2024 & 16kb Payload & Quest to Quest Wireless & 115 ms & 60 ms & 200 ms \\ \hline
    14-12-2024 & 16kb Payload & Quest to Quest Hotspot & 90 ms & 220 ms & 195 ms \\ \hline

    \end{tabular}
    }
    \caption{Average latency measurements for different connection types, scenarios, and configurations across all test scenarios.}
    \label{fig:latency-measurements-expanded}
\end{figure}

\begin{figure}[H]
    \includegraphics[width=0.42\textwidth]{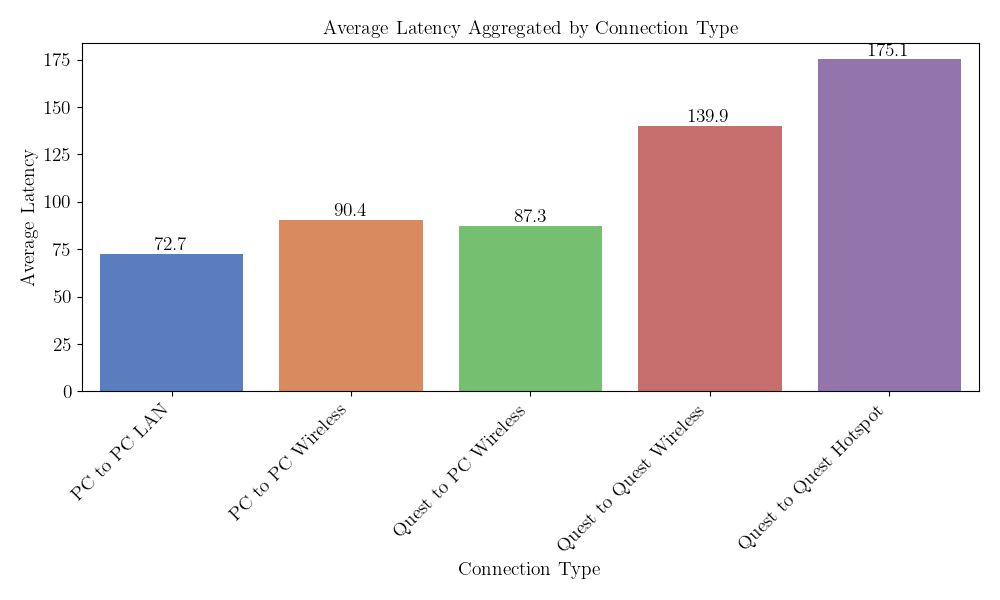}
    \caption{Average Latency Across Connection Types: PC to PC LAN shows the lowest latency, while Quest to Quest Hotspot exhibits the highest.}
    \label{fig:example2}
\end{figure}

\begin{figure}[H]
    \includegraphics[width=0.42\textwidth]{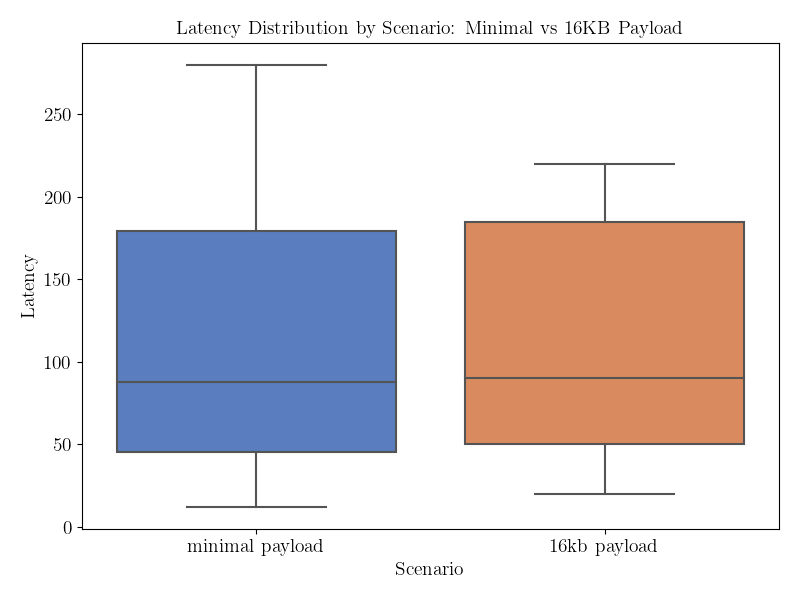}
    \caption{Latency Distribution Across Scenarios with Outliers: Minimal payload shows slightly lower median latency compared to 16KB payload. A larger sample is necessary to understand trends, as WebSocket's message prioritization might influence results.}
    \label{fig:example3}
\end{figure}

\begin{figure}[H]
    \includegraphics[width=0.42\textwidth]{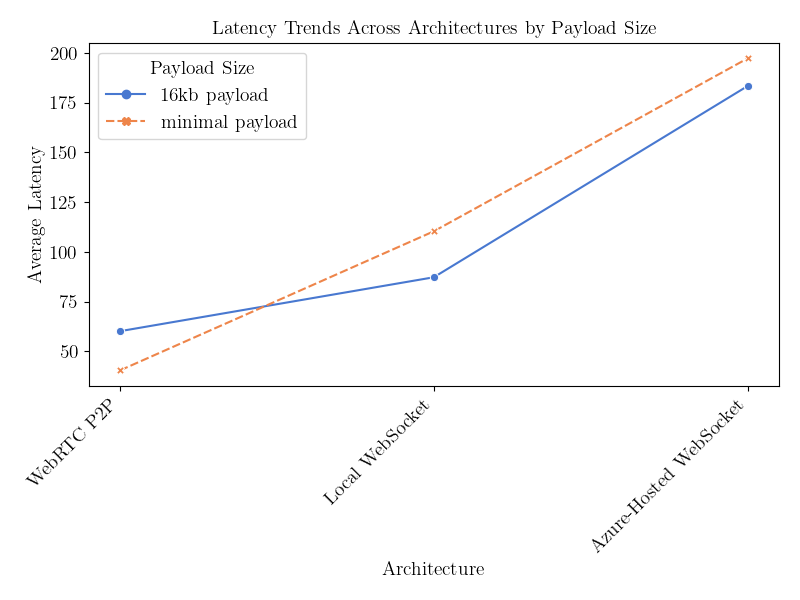}
    \caption{Latency Trends by Architecture and Payload Size: Minimal payload is only faster on WebRTC P2P, while 16KB payload performs better on WebSocket architectures.}
    \label{fig:example4}  
\end{figure} 

\begin{figure*}[ht]
    \rotatebox{-90}{
        \includegraphics[width=1.2\textwidth]{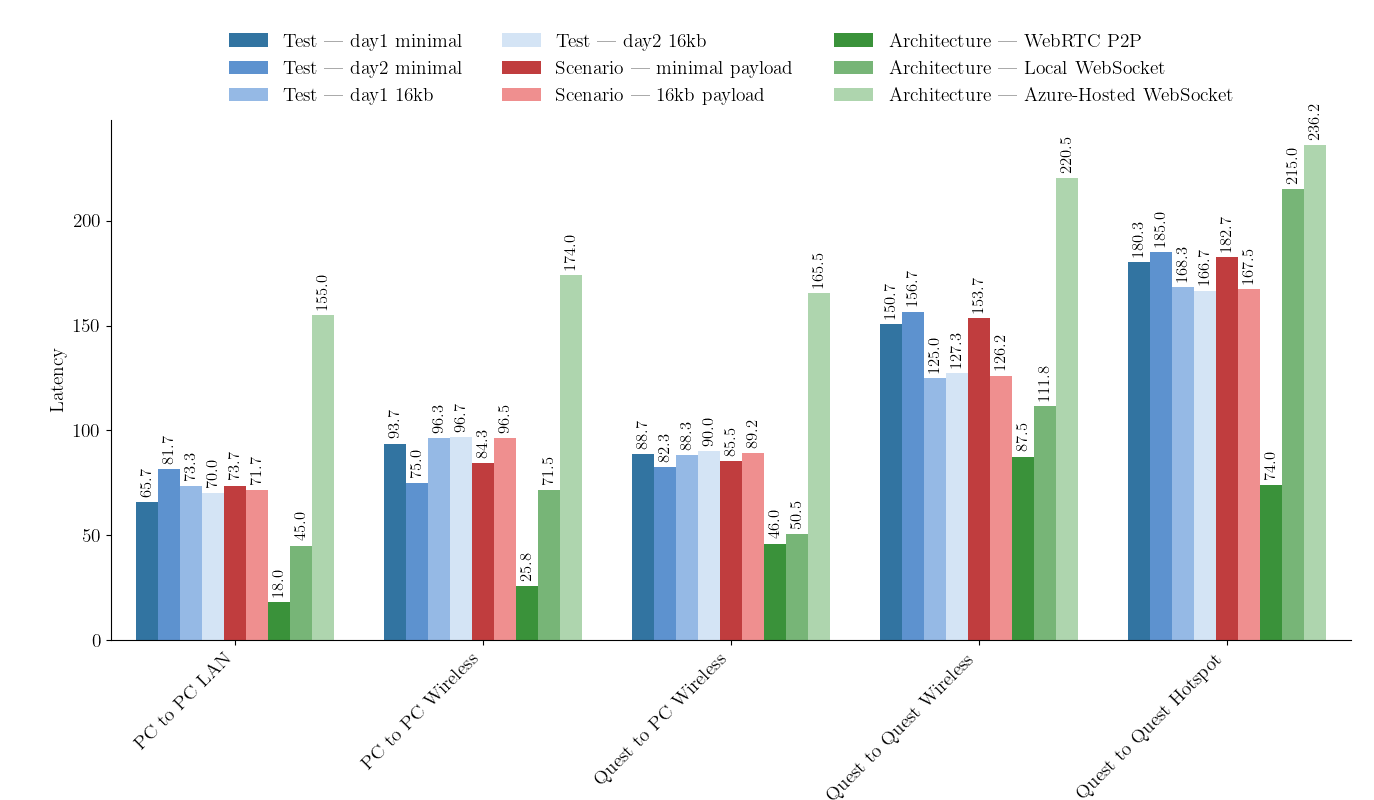}
    }
    \caption{Breakdown across connection types, test scenarios, payload sizes, and architectures.} 
    \label{fig:example1}
\end{figure*}

\begin{table}[H]
\centering
\begin{minipage}{0.4\textwidth}
\raggedright
\textbf{Networks}
\begin{enumerate}
    \item Commercial office network (Vodafone 5GHz, 500Mbps)
    \item Mobile hotspot (NOS 5G, Xiaomi Mi 11T)
\end{enumerate}

\textbf{Devices}
\begin{enumerate}
    \item PC (Windows 11, Ryzen 5 5500, 32GB RAM)
    \item MacBook Air (macOS Sonoma, M1, 8GB RAM)
    \item Meta Quest 3 (Android 11, Snapdragon XR2, 8GB RAM)
\end{enumerate}
\end{minipage}
\caption{Listing of networks and devices used in the experiment.}
\label{tab:networks_devices}
\end{table}

\vspace{3cm}

\end{document}